\documentclass[aps,superscriptaddress,preprintnumbers,nofootinbib,showkeys]{revtex4-1}
\usepackage{graphicx,amssymb,amsmath,color}
\usepackage{hyperref}

\newcommand{\beq}{\begin{equation}}
\newcommand{\eeq}{\end{equation}}
\def\bea{\begin{eqnarray}}
\def\eea{\end{eqnarray}}
\newcommand{\bei}{\begin{itemize}}
\newcommand{\eei}{\end{itemize}}
\newcommand{\bmat}{\begin{matrix}}
\newcommand{\emat}{\end{matrix}}

\newcommand{\Fig}[1]{Fig.~\ref{#1}}
\newcommand{\Eq}[1]{Eq.~(\ref{#1})}
\newcommand{\Sec}[1]{Sec.~\ref{#1}}

\def\={\,=\,}
\def\+{\,+\,}
\def\-{\,-\,}
\def\mDM{m_{\rm DM}}
\def\Mch{{\cal M}}
\def\Msun{M_\odot}

\begin{document}

\title{A New Probe of Dark Matter-Induced Fifth Force \\with Neutron Star Inspirals}

\author{Han Gil Choi}
\email{alivespace@snu.ac.kr}
\affiliation{Center for Theoretical Physics, Department of Physics and Astronomy, \\Seoul National University, Seoul 08826, Korea}

\author{Sunghoon Jung}
\email{sunghoonj@snu.ac.kr}
\affiliation{Center for Theoretical Physics, Department of Physics and Astronomy, \\Seoul National University, Seoul 08826, Korea}


\begin{abstract}

A light scalar dark matter (DM) is allowed in a wide range of its mass and interaction types. We show that the light scalar DM may be probed in a new way from final years of neutron-star (NS) binary inspirals. If the DM interacts with the neutron, its long wave coherence in the background can induce the time-oscillating mass shift, to which the binary inspiral is inherently sensitive. But the sensitivity is found to be significantly enhanced by a large number of gravitational-wave (GW) cycles during year-long highest-frequency measurements in the broadband $f \simeq 0.01-1000$ Hz. Such broadband measurements that can be realized by a future detector network including LIGO and mid-band detectors can probe unconstrained parameter space of the light scalar DM.

\end{abstract}


\maketitle

\tableofcontents

\section{Introduction}

The Gravitational Wave (GW) from compact binary mergers are finally discovered~\cite{Abbott:2016blz}. The discovery has convinced the existence of solar-mass black holes for the first time~\cite{TheLIGOScientific:2016htt} and tested General Relativity in a new way~\cite{TheLIGOScientific:2016src}. But more and foremost excitingly, in the upcoming years with next-generation GW detectors, its physics potential is bound to grow significantly and extend outside astrophysics. 

In particular, binary neutron stars (NS)~\cite{TheLIGOScientific:2017qsa} are becoming new precision observatories. Their final years of inspirals can be tracked very precisely through the observation of GW radiation. The final inspiral is strongly governed by general relativity, producing a well-predicted particular type of evolution called the ``chirping''. The chirping inspiral is thus naturally immune to small perturbations from their environments or backgrounds. This allows not only the detection of binary GWs as tiny as $10^{-21}$ fractional oscillation of the metric but also precision cosmology combined with optical counterparts~\cite{Abbott:2017xzu}.

But the immunity does not mean that small perturbations are irrelevant or unobservable. Often, they do leave important traces on the binary inspiral from which we can observe the binary environments -- the Universe. One of the most exciting environmental effects would be from the dark matter (DM). For example, DM can accumulate at the core of NS, strongly modifying NS binary mergers~\cite{Nelson:2018xtr,Croon:2017zcu,Ellis:2017jgp,Ellis:2018bkr}. A DM locus nearby the binary may be also able to perturb the binary orbit in such a way to enhance the instability or ellipticity~\cite{Randall:2017jop}. The high-precision observation of NS inspirals with GW may have broader (unexplored) sensitivities to milder DM effects from more varieties of DM. 

Dark matter is one of the biggest mysteries of the Universe. In spite of its unprecedented evidence, it has not been discovered yet. For decades, Weakly Interacting Massive Particle -- WIMP -- has been a main paradigm of DM, but all DM detection experiments so far have failed to discover it. Today, it became essential to explore broader possibilities of DM interactions and masses both theoretically and experimentally.

A light scalar DM is one important candidate that receives much attention these days. A wide range of its mass as light as $10^{-23}$ eV is unconstrained. Various types of its couplings to matter are possible. There are also several well-motivated examples such as axions~\cite{Kim:2008hd}, fuzzy dark matter~\cite{Hu:2000ke}, relaxions~\cite{Graham:2015cka} as well as simple scalar DM. These scalars might be relevant to the solution of important particle physics problems such as strong QCD problem and the Planck-weak hierarchy. Thus, many direct detection experiments are proposed and ongoing; a good survey of them is in Ref.~\cite{Graham:2015ifn}. But to probe a complete range of possible masses and interactions of the DM, more new and complementary ideas are needed.

In this paper, we show that final years of NS-NS inspirals may be sensitive to light scalar DM-induced perturbations in a new way.
We give an overview of the new observable and possible experimental setup in \Sec{sec:overview}, then we introduce DM models in \Sec{sec:model}, discuss new observables and other existing ones in \Sec{sec:signal}, experimental setup and calculation in \Sec{sec:setup}, finally show and analyze results in \Sec{sec:prospects} and \ref{sec:analysis}, and conclude at the end.

\section{Overview} \label{sec:overview}

We give an overview of the new DM-induced signal on the NS-NS inspiral, a signal that can be observed through the last years of inspiral at, e.g., LIGO plus mid-frequency detectors. 
\bei
\item Signal with the oscillating NS mass. The light scalar DM (not just light scalars) interacting with neutrons can induce the \emph{time-oscillating mass-shift} of NS-NS binaries. The oscillation is due to the lightness of DM with long phase coherence. The phase coherence is kept for long enough periods $\sim 1/\mDM v^2 \gg 1/\mDM$ because DM is non-relativistic $v\sim10^{-3}$. Thus, the lightest possible DM oscillates coherently at its Compton frequency, $\mDM \, \gtrsim \, 10^{-22} \, {\rm eV} \approx 0.76$/year, which is about 1 per year.
\item Enhanced sensitivity to the chirp-mass. The oscillation in time is a key property that provides a time-dependent change to detect the mass shift.  As long as the DM Compton period is within inspiral measurement time, the oscillation is detectable, through the exquisite chirp-mass measurement from GW waveform evolution. The sensitivity benefits from \emph{a large number of GW cycles} during a long measurement, which can accumulate a tiny phase shift (from the chirp-mass shift) in each cycle to a detectably large one. Thus, the chirp-mass accuracy is augmented by $\sim N_{\rm cyc} \cdot {\rm SNR}$ (not just SNR). 
\item Highest-frequency broadband. The broadband $f \simeq 0.01-1000$ Hz (e.g. with LIGO + mid-frequency detectors) is ideal to detect the DM effects, as it is the \emph{highest-frequency band with year-long lifetime} of NS-NS binaries. Not only can a long measurement enhance $N_{\rm cyc}$ and SNR, but the highest-frequency end can also resolve important parameter degeneracies, partly by utilizing the Doppler effects around the Sun. In addition, as mentioned, the year-long measurement is also a proper time scale to probe the lightest scalar DM.
\item NS-NS. The NS-NS is the type of binaries that can test the DM induced effect. Here, non-DM light-scalar effects (such as dipole radiation of the scalar and Yukawa force) is absent or at least suppressed. We focus on the NS-NS case in this paper. 
\item Other probes. There are existing and proposed experiments that can be sensitive to light-scalar DM effects: pulsar timing arrays, lunar laser ranging, atomic clocks, GW interferometers, and torsion balance experiments. These can search for the DM-induced time-varying clock rate, $\alpha$, acceleration, and equivalence-principle (EP) violation. Our new probe -- looking for the DM-induced absolute mass-shift -- can be competitive or complementary to those. 
\eei

\section{Models of DM-induced fifth force} \label{sec:model}

A light scalar DM shows its wave nature through its long phase coherence. Although background DM is an incoherent superposition of individual DM waves, their phase coherence is retained for a long time $\sim 1/(m_{\rm DM} v^2) \gg 1/m_{\rm DM}$. Within that time, DM then coherently oscillates at the Compton frequency $\mDM = 2.42 \times 10^{-8} {\rm Hz} \left( \frac{ \mDM}{10^{-22} {\rm eV}} \right)$, and its background effect can be collectively enhanced.

Testable signals arise when the light scalar DM interacts with the visible matter (hence, the fifth force), in particular with the neutron in this paper. We introduce benchmark models for this phenomena: 
\bei 
\item Higgs portal DM. The mixing between the DM $\phi$ and the Higgs $h$ induces the coupling to the neutron $n$
\beq 
{\cal L} \, \supset \,  \frac{b \phi}{m_h^2} \langle h \rangle g_{hnn} \bar{n} n  \quad \to \quad \frac{b \phi}{m_h^2} c_N m_n \bar{n} n,
\eeq
where the non-perturbative QCD effects on the nucleon coupling is captured by $c_N \approx 200 \, {\rm MeV} / m_n$ with significant uncertainties~\cite{Piazza:2010ye}. The DM wave almost coherently oscillates in time and space with the amplitude set by the DM density: $\phi(t) = (\sqrt{\rho_{\rm DM}} / m_\phi) \, \cos ( m_\phi t )$. The coefficient $b$ is our free parameter. 

The couplings to protons and electrons are also generated by the mixing; since these couplings are not proportional to the masses due partly to QCD confinement effect, they generally break the (weak) equivalence principle (EP)~\cite{Piazza:2010ye}. For constraints on the weak EP-violation, we take the results in Ref.~\cite{Graham:2015ifn}. Couplings to photons and gluons can also be generated (at least through loops of charged/colored particles). But searches of such couplings from the lightest possible DM is absent. We focus on the coupling to the neutron in this paper.

%
\item Scalar DM coupled to the trace of the stress-energy tensor $T=T^{\mu}_{\mu}$ 
\beq
{\cal L}  \, \supset \, g_{\phi}\phi T ,
\eeq
where $g_{\phi}$ is a universal coupling constant. In the long-wavelength limit of $\phi$, the interaction term for the neutron is effectively equivalent to
\beq
{\cal L}  \, \supset \, g_{\phi n n} \, \phi \bar{n} n,
\eeq
where $g_{\phi n n}=g_{\phi} m_n$ is our free parameter specialized to the neutrons. We assume no other modifications in the gravitational sector. This model does not violate the weak EP, but the strong EP is still broken. One can find that $\phi$ violates the strong EP in the following two ways: (1) the outcome of the elementary particle mass measurements depend on the space-time varying intensity of $\phi$, (2) the self-energy dependence (due to the self-energy of $\phi$) of the free-falling acceleration under the external fields ($g_{\mu \nu}$ and $\phi$).
The effect of the scalar field $\phi$ can be constrained by not only the tests of the fifth force but also by the tests of the general relativity, such as the Shapiro delay measurement from Cassini spacecraft~\cite{Bertotti:2003rm} and strong-EP tests by the observation of stellar binary or triple systems containing a pulsar~\cite{Freire:2012nb,Archibald:2018oxs}. The Cassini constraints are stronger and will be shown in our results.
\eei

\section{Signals in Neutron Star inspirals} \label{sec:signal}

We introduce the new observable in \Sec{sec:signal1}, then we review non-DM signals in NS-NS in \Sec{sec:signal2}, other mass-shift effects in \Sec{sec:signal3}, other light scalar non-DM effect \Sec{sec:signal4} and equivalence-principle violating effects in \Sec{sec:signal5}.

\subsection{Oscillating NS-NS mass-shift} \label{sec:signal1}

The neutron-star(NS) inspiral interacts with the background DM distributed over the space through which it moves. This interaction can leave traces\footnote{Similar DM effects on binary pulsars have been studied in Ref.~\cite{Blas:2016ddr} based on DM oscillation in resonant with binary orbital frequency.} as (1) the oscillating mass-shift of NS, (2) oscillating external forces on the NS.
The former one is our focus in this paper, and we show that NS-NS inspirals can be used to test this DM effect.

The (oscillating) fractional neutron mass-shift from the two benchmark models is
\beq
\frac{\delta m_n}{m_n} \= \left\{ \bmat  
c_N \frac{b \phi}{m_h^2} \= c_N \frac{b \sqrt{\rho_{\rm DM}}}{m_\phi m_h^2} \,\simeq \, 8.0 \times 10^{-13} \left( \frac{b}{10^{-9} {\rm eV}} \right) \left( \frac{10^{-7} {\rm Hz}}{m_\phi} \right)  \, \cos(m_\phi t)\\
\frac{g_{\phi nn}  \phi}{m_n} \= \frac{g_{\phi nn} \sqrt{\rho_{DM}}}{m_\phi m_n} \, \simeq \, 6.3 \times 10^{-13} \left( \frac{g_{\phi nn}}{10^{-23}} \right) \left( \frac{10^{-7} {\rm Hz}}{m_\phi} \right) \, \cos(m_\phi t)
\emat \right.
\eeq
The effect is proportional to $\phi \propto \sqrt{\rho_{\rm DM}}$ so that galactic centers where the majority of both DM and NS-NS reside are good places to detect the DM effect. We base our numerical calculation on the value of $\rho_{\rm DM} \= 77.3 \, {\rm GeV/cm}^3$ from the 0.8 kpc flat-core value of Burkert profile, but we also consider variations later. 
The neutron mass-shift will induce the NS mass-shift, hence the NS-NS chirp-mass shift
\beq
\frac{\delta \Mch}{\Mch} \= a \frac{\delta m_n}{m_n} 
\label{eq:mass-shift} \eeq
with $a= 1$ for NS-NS binary, but there can be a mild suppression from the neutron fraction in the NS. 

The mass-shift becomes observable as it oscillates in time: $\delta \Mch(t) / \Mch  \propto \sqrt{\rho_{\rm DM}} \cos (m_\phi t)$. The time-oscillation of the chirp mass induces a characteristic change of the GW evolution that cannot be mimicked by GR effects. As a proxy of sensitivity, we will calculate the parameter space where the oscillation amplitude equals to the chirp-mass measurement accuracy; we discuss our calculation in the next section and show results in \Fig{fig:single10} and \ref{fig:allNSNS}. The chirp mass can be exquisitely well measured through a huge number of GW cycles and highest-frequency data, as will be discussed. For the chirp-mass oscillation to be detected, at least a large portion of an oscillation should be within the GW measurement time (about a year or longer in $f \geq 0.01$ Hz). Since $m_\phi \gtrsim 10^{-22} \, {\rm eV} = 0.76/{\rm yr}$ and their phase coherence is retained for much longer time, the (multi) year-long high-frequency GW measurement is proper to test the lightest possible DM.

\subsection{Other light-scalar (non-DM) effects in NS inspirals} \label{sec:signal2}

A light scalar can also induce other effects in NS-NS binaries, the non-DM effects that exist even if the scalar is not the main fraction of DM. 
The exchange of light scalars $\phi$ mediates a long-range Yukawa force between the neutron stars, deviating from the $1/r^2$ law
\beq
\mu \frac{v^2}{r} \= \frac{G \mu M}{r^2} \left( 1 + \alpha(1 + m_\phi r ) e^{-m_\phi r} \right),
\eeq
where $\alpha \= \frac{b^2 c_N^2}{m_h^4} \frac{1 }{4\pi G}\, \simeq \, 1.48 \times 10^{-9} ( \frac{b}{ 10^{-9}\, {\rm eV} })^2$ for the first model and
$\alpha  \= \frac{g_{\phi nn}^2 }{4\pi G m_n^2} \, \simeq \, 1.44 \times 10^{-9} ( \frac{g_{\phi nn}}{10^{-23}})^2$ for the second model, and the reduced and total mass $\mu$ and $M$. 
The effect on the GW waveform evolution can be described approximately by the shift of the chirp mass 
$\frac{ \delta \Mch (r)}{\Mch}  \, \simeq \, \frac{2}{5} \alpha \, (1+ m_\phi r) \, e^{-m_\phi r}$\footnote{Our method applied to ET yields similar or actually slightly worse sensitivity than the more dedicated estimation in Ref.~\cite{Alexander:2018qzg}.}. 
The resulting radius-dependent (hence, frequency-dependent) chirp mass is a clean signal that cannot be mimicked by GR effects. The total change of the chirp mass during a measurement starting from $f_i$ (or $r_i$) until $f_f$ (or $r_f$) is given by
\beq
\frac{\delta \Mch}{\Mch} \, \simeq \, \frac{2}{5}\alpha  \Big( (1+m_\phi r_f) \, e^{-m_\phi r_f} - (1+m_\phi r_i)\, e^{-m_\phi r_i} \Big) \, \approx \, \left\{ \bmat \frac{2}{5}\alpha (1- \frac{1}{2}m_\phi^2 r_f^2), &  r_i \gg \frac{1}{m_\phi}  \\  \frac{1}{5}\alpha m_\phi^2 (r_i^2-r_f^2), &  r_i \ll \frac{1}{m_\phi}  \emat \right.  \, \,  \lesssim \, \, \frac{2}{5} \alpha
\label{eq:yukawa} \eeq
The change is maximal, $\frac{2}{5} \alpha$, for the scalar mass in the range $r_i \gg 1/m_\phi \gg r_f$.
The range, for the NS-NS binary, is $r_i(f=0.1 \, {\rm Hz}) \simeq (7.8 \times 10^{-14} \,{\rm eV})^{-1} \simeq 16000 \, {\rm km}$ and $r_f(f=1000 \, {\rm Hz}) \simeq (3.7 \times 10^{-11} \, {\rm eV})^{-1} \simeq 34 \, {\rm km}$\footnote{For LIGO-band expected sensitivities, refer to Refs.~\cite{Croon:2017zcu,Huang:2018pbu,Kopp:2018jom,Alexander:2018qzg}.}. Thus, this effect can only probe those range of the DM mass (in our experimental setups); lighter Yukawa force essentially looks the same as gravity. In \Fig{fig:single10} and \ref{fig:allNSNS}, we show the parameter space where \Eq{eq:yukawa} equals to the chirp-mass accuracy. 


\medskip
A light scalar (again not necessarily the main DM) can also be efficiently radiated if each NS carries different scalar charge-to-mass ratio, forming a scalar-charge dipole~\cite{Croon:2017zcu, Huang:2018pbu}. This dipole radiation is qualitatively different from the GW quadrupole radiation, thus can be tested with GW waveform evolution~\cite{Huang:2018pbu,Kopp:2018jom,Alexander:2018qzg}. It is efficient for any light scalars with long enough Compton wavelength $1/m_\phi \gtrsim 10$ km. But this effect is absent in the NS-NS in our model, as every NS has the same charge ($m_{\rm NS}/m_n$) to mass ($m_{\rm NS}$) ratio $=1/m_n$; at least, the radiation is suppressed by a small variation of the neutron fraction in the NS. In the NS-BH, on the other hand, the dipole radiation is efficient and dominant effect of light scalars, prohibiting the detection of the DM effects -- the oscillating mass-shift. Thus, in this paper, we focus on the NS-NS as the type of binaries that can test the light-scalar DM effects.

\subsection{Mass-shift in other experiments} \label{sec:signal3}

Pulsars are highly stable and regular clocks. If its mass changes by DM effects, its rotation period (hence, the clock) also changes; see also Ref.~\cite{Blas:2016ddr}. This leaves an oscillating timing residual on each pulsar timing measurement. Each pulsar's variation is uncorrelated with those of every other pulsar since pulsars are separated by distances much longer than the DM Compton wavelength. Thus, the average of pulsar timing array (PTA) can provide a stable clock, cancelling the DM effect~\cite{Graham:2015ifn}. This PTA clock may be compared with individual pulsar timing to measure the oscillation. 

With EP violations, the pulsar measurement can also be affected by the variation of atomic clocks on the Earth~\cite{Graham:2015ifn}. The atomic-clock rate varies because atomic transition frequencies are affected by the DM-induced variations of nucleus (and electron) masses and fine-structure constant $\alpha$~\cite{Arvanitaki:2015iga,VanTilburg:2015oza,Arvanitaki:2014faa,Stadnik:2015xbn,Stadnik:2014tta}. But all the atomic clocks oscillate coherently on the Earth while independently from all the pulsars. Thus, individual pulsar's oscillation can perhaps be distinguished from atomic clock oscillations. 

Therefore, we assume that DM-induced mass-shift can be detected by either observation, whether EP is conserved or not. One difference is that DM density at pulsars can be different from local density that affects the atomic clocks on the Earth. In the future, this difference can be exploited to better measure DM effects. Today, however, most pulsars used in IPTA~\cite{Verbiest:2016vem} and Parkes PTA~\cite{Porayko:2018sfa} are within $1 \sim 2$ kpc from the Earth\footnote{Presumably, pulsars and NS-NS are both accumulated at galactic center (GC). But the observed distributions (pulsars with lights and NS-NS with GWs) could be somewhat different. See Sec.~\ref{sec:transparent} for more discussions.}. Thus, we use the same local DM density $\rho_{\rm DM} = 0.39 \, {\rm GeV}/{\rm cm}^3$ to estimate both effects. They give equivalent sensitivities at the end so that we essentially do not distinguish the two observables. 

The first release of IPTA~\cite{Verbiest:2016vem} achieved the timing sensitivity $\Delta t \, \approx\, \sigma_t/\sqrt{ N_p N_m}$ with r.m.s. timing residual $\sigma_t \simeq 1 \, \mu {\rm s}$, $N_p=50$ pulsars and $N_m \simeq 10 \, {\rm yr}/ 2 \,{\rm weeks}$ total number of measurements.  In our final results, we compare this sensitivity with the DM-induced mass-shift or timing-residual amplitude
\beq
\Delta t \simeq  \int dt \, \frac{\delta T }{ T}  \simeq  \frac{1}{m_\phi} \frac{\delta m_n }{ m_n}.
\eeq

Lunar laser ranging (LLR) may be also sensitive to the absolute mass-shifts of the Earth and the Moon. About 50\% of their masses come from neutrons, so their fractional mass-shift is approximately $\frac{1}{2} \frac{\delta m_n }{ m_n}$. We assume that the mass-shift induces a change or perturbation in orbital radius about the same fractional size as the fractional mass-shift. The LLR measurement of the separation distance is currently limited by $\delta \ell \leq 1 \sim 2$ cm~\cite{Williams:2012nc}. We take the fractional sensitivity on the mass-shift to be 
\beq
\frac{ \delta \ell }{ (3.8 \times 10^{5} \, {\rm km}) } \, \simeq  \frac{1}{2} \frac{ \delta m_n }{ m_n}\, \lesssim \, 5 \times 10^{-11} .
\eeq
In our final results, we compare this sensitivity with the DM-induced mass-shift.

It is also proposed that GW interferometers can detect DM-induced space-time varying accelerations on the mirrors~\cite{Graham:2015ifn,Arvanitaki:2016fyj,Pierce:2018xmy}. The best sensitivity is achieved when the DM Compton frequency matches approximately with the interferometer sensitivity range, hence $m_\phi \sim $ 100 Hz or $10^{-3}$ Hz for LIGO and LISA, for example.

\subsection{Light-scalar(non-DM) effects in other experiments}\label{sec:signal4}

A long-range Yukawa force has been also searched in other observations: lunar laser ranging (LLR)~\cite{Williams:2012nc} and Keplerian tests from planetary motions. We show these existing constraints taken from Ref.~\cite{Adelberger:2003zx}. Note that the searches only cover up to 1 AU($\sim 10^{-18}$ eV) scale. 

In contrast, the Shapiro time delay measurement of the Cassini spacecraft~\cite{Bertotti:2003rm} can give constraints above 1 AU too. During the period of 15 days before and after the Cassini solar conjunction event, the Cassini spacecraft and the ground antenna on the Earth have exchanged the radio signals. The strong gravity of the Sun delays the round-trip time of the radio signal by $\Delta t$
\beq
    \Delta t \, \propto \, 4\frac{G M_\odot}{c^3} \=  4(1- \alpha) \frac{G M_\odot^{\rm orbital}}{c^3},
\label{eq:Shapiro1}\eeq
where $M_{\odot}^{\text{orbital}}=M_{\odot}(1+\alpha)$ is what is determined by the orbital motion (under the influence of both gravity and the $\phi$-Yukawa force). The time-delay of the radio siganl is not affected by the $\phi$-Yukawa force.
Thus, the $\phi$-Yukawa force (or, the strong EP-violation) can be searched by comparing the two effects as in \Eq{eq:Shapiro1}.
$2\alpha \lesssim 2.1 \times 10^{-5}$ from Cassini experiment~\cite{Bertotti:2003rm}. This is weaker than the static EP-violation searches for the first model, but is relevant to the strong EP-violating second model. Other constraints from stellar binaries or triple systems containing a pulsar~\cite{Freire:2012nb,Archibald:2018oxs} are weaker.

\subsection{Equivalence-principle violation} \label{sec:signal5}

There are existing and proposed experiments that can be sensitive to light-scalar EP violation. Stringent limits were obtained from the EP tests of E\"{o}t-Wash torsion balance~\cite{Wagner:2012ui} and MICROSCOPE free-falling Earth orbit~\cite{Touboul:2017grn} experiments (see also Ref.~\cite{Hees:2018fpg}). But both measure static non-DM effects. We show the E\"{o}t-Wash constraint taken from Ref.~\cite{Graham:2015ifn} in our final results. 

On the other hand, atomic clocks can be sensitive to EP-violating DM effects. Depending on the atom's proton and neutron fractions, oscillating DM can induce different variations of clock rate. To measure the differences, one can compare clock rates among atomic clocks made of different atoms~\cite{Arvanitaki:2015iga,VanTilburg:2015oza,Arvanitaki:2014faa,Hees:2016gop,Stadnik:2016zkf}, or accelerations among atom-interferometers made of different atoms~\cite{Graham:2015ifn}, or the PTA clock rates measured by different atomic clocks~\cite{Graham:2015ifn}. Torsion balance experiments may also be sensitive to DM effects by sensing DM-induced force directions that may not point to the Earth. These searches all depend on the local DM oscillation. We show existing constraints from atomic clock experiments~\cite{VanTilburg:2015oza,Hees:2016gop} in our final results.

\section{Broadband GW detectors and Calculation} \label{sec:setup}

We discuss proposed experimental setups and our calculation.

The crucial for exquisite chirp-mass measurement are a large number of GW cycles and highest-frequency chirping. Thus we consider final 1-year and 10-year measurements of NS-NS binaries, long enough and highest-frequency measurements that are proper for this work. At 1$\sim$10 years before the merger, the redshifted GW frequencies are ${\cal O}(0.01-0.1)$ Hz (where the lower range corresponds to $z_S\simeq 10$) and reach the innermost stable orbit (ISCO) at ${\cal O}(100-1000)$ Hz. We combine mid- and high-frequency detectors to cover those range of frequencies $0.01 \, {\rm Hz} \lesssim f \lesssim 1000$ Hz.  

The first benchmark detector-network consists of 4 sets of aLIGO (design sensitivity)~\cite{TheLIGOScientific:2016agk} + Atom Interferometer (AI) (resonant mode)~\cite{Graham:2016plp,Graham:2017pmn}; and the more optimistic benchmark network consists of one set of Einstien Telescope (ET)~\cite{Hild:2010id} + Big Bang Observatory (BBO)~\cite{Yagi:2011wg}. The second benchmark has ${\cal O}(10)$ times smaller noise. 

Our calculation proceeds as follow. Since we consider year-long or longer measurements, particular detector properties (baseline direction and their rotation, etc) are not so important. Thus, for the calculational simplicity, we use the simplest antenna function (from a single-baseline AI detector) for all kinds of detectors mentioned above; we follow the procedure in Ref.~\cite{Graham:2017lmg}. Considering more accurate and complicated antenna functions may even improve the chirp-mass precision, thus our estimation may be conservative in this sense. For GW waveforms, we use $m_{\rm NS} = 1.3 \Msun$ 
with the amplitude at the Newtonian order. The GW phase includes post-Newtonian corrections up to 1.5PN at which the reduced mass $\mu$ and spin-orbit parameter $\beta$. We pick a random set of extrinsic source parameters (sky location, polarization, orbit inclination) that are close to the orientation-averaged GW amplitude; we use the same parameters as in Ref.~\cite{Graham:2017lmg}. For further simplicity, we assume that spins and orbital eccentricities are zero.

Our goal is to compare the DM-induced chirp-mass shift in \Eq{eq:mass-shift} with the chirp-mass measurement accuracy, as a proxy of sensitivity to the DM effect. We envisage that oscillating data have the sensitivity to the oscillating part of the chirp mass at the same level as the chirp-mass accuracy without oscillations; thus, inspirals can be sensitive equally to any oscillation frequency as long as it is within the inspiral measurement time. The chirp-mass accuracy (without oscillations) is estimated by the Fisher information matrix, $F$. We include 10 source parameters: sky location ($\theta, \phi$), polarization, inclination, luminosity distance $D_L$, coalescence time $t_c$, constant phase-shift, masses (chirp mass $\Mch$ and reduced mass $\mu$), and spin-orbit coupling $\beta$ (although we set spins to zero, we do not assume that we know it). The parameter definition and calculation are followed from Refs.~\cite{Cutler:1997ta,Graham:2017lmg}. Fisher matrices from different detectors are added linearly; equivalently, SNR is added in quadrature. The accuracy of a parameter is given by the square-root the inverse-Fisher diagonal elements $\sqrt{(F^{-1})_{ii}}$; thus, it improves with the square root of the number of measurements and linearly with SNR.

\section{Prospects} \label{sec:prospects}

\begin{figure}[t] \centering
\includegraphics[width=0.49\textwidth]{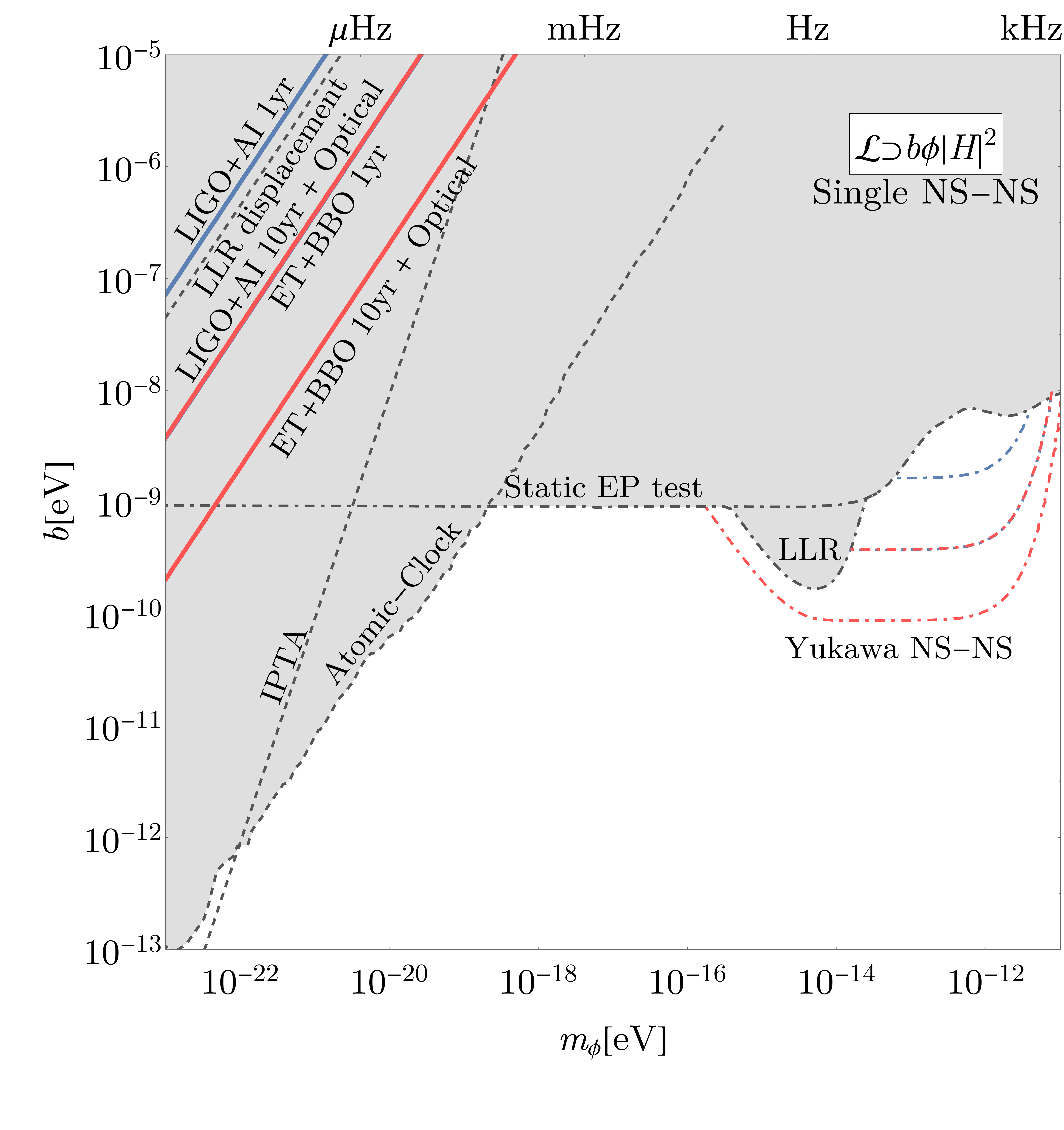}
\includegraphics[width=0.49\textwidth]{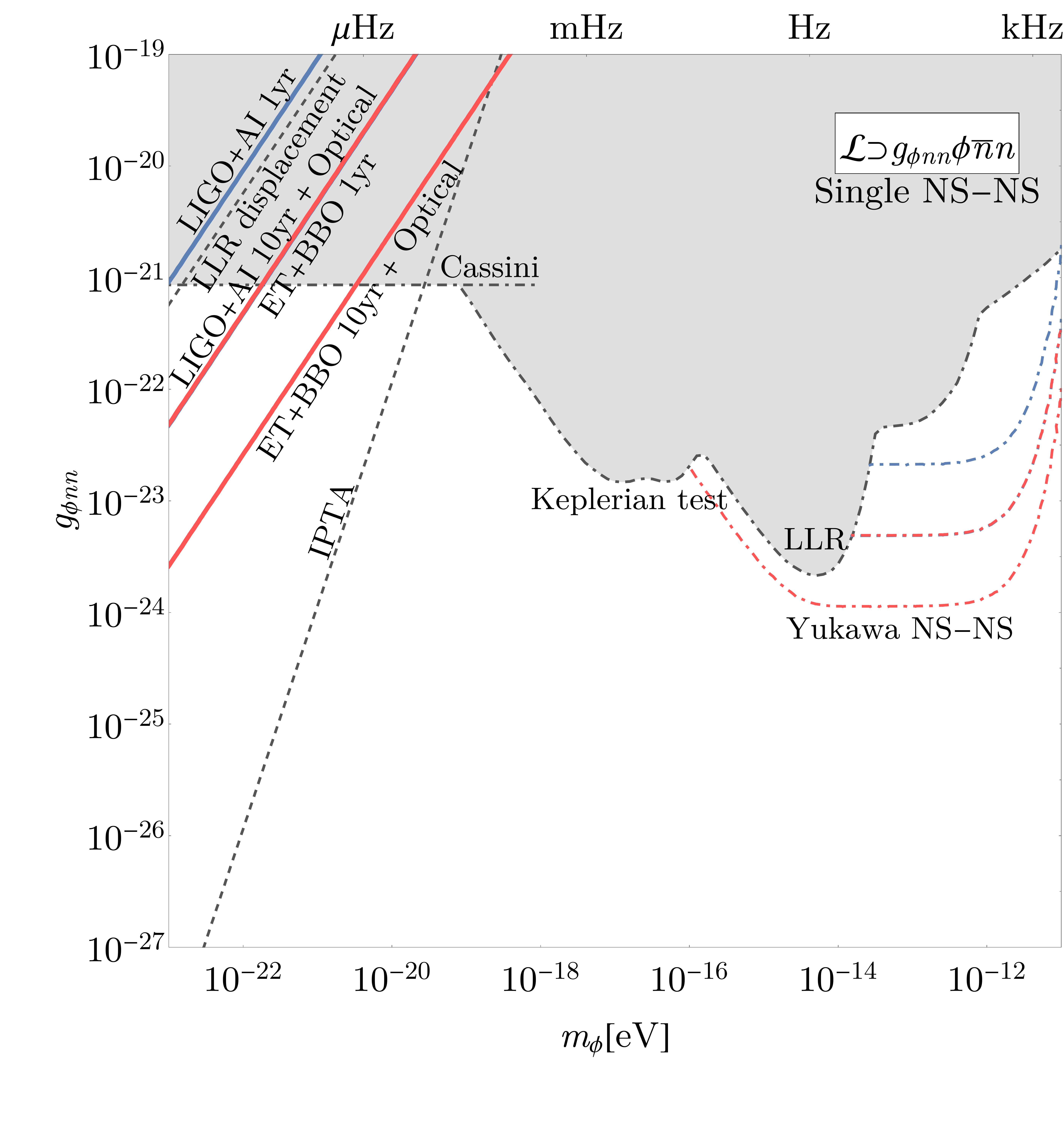}
\caption{Single measurement of NS-NS at 10 Mpc with 4(aLIGO+AI) (blue) or with ET+BBO (red). Along the shown solid lines, the DM-induced mass-shift $\delta \Mch/ \Mch$ equals to the chirp-mass accuracy. Each solid line corresponds to different setup and assumptions; see text for details. The aLIGO+AI 10yr and ET+BBO 1yr curves overlap. Higgs-portal model (left) and scalar-coupled to the trace of the stress-energy tensor (right). Also shown are existing (shaded) and reinterpreted (non-shaded) constraints on DM-induced effects (dashed) from IPTA, LLR displacement, and atomic clocks~\cite{VanTilburg:2015oza,Hees:2016gop} and on non-DM effects (dot-dashed) from static EP test~\cite{Wagner:2012ui,Graham:2015ifn} and Yukawa searches with LLR~\cite{Williams:2012nc}, Keplerian test~\cite{Adelberger:2003zx}, Cassini~\cite{Bertotti:2003rm}, and NS-NS inspiral.}
\label{fig:single10}
\end{figure}
\begin{figure}[t] \centering
\includegraphics[width=0.49\textwidth]{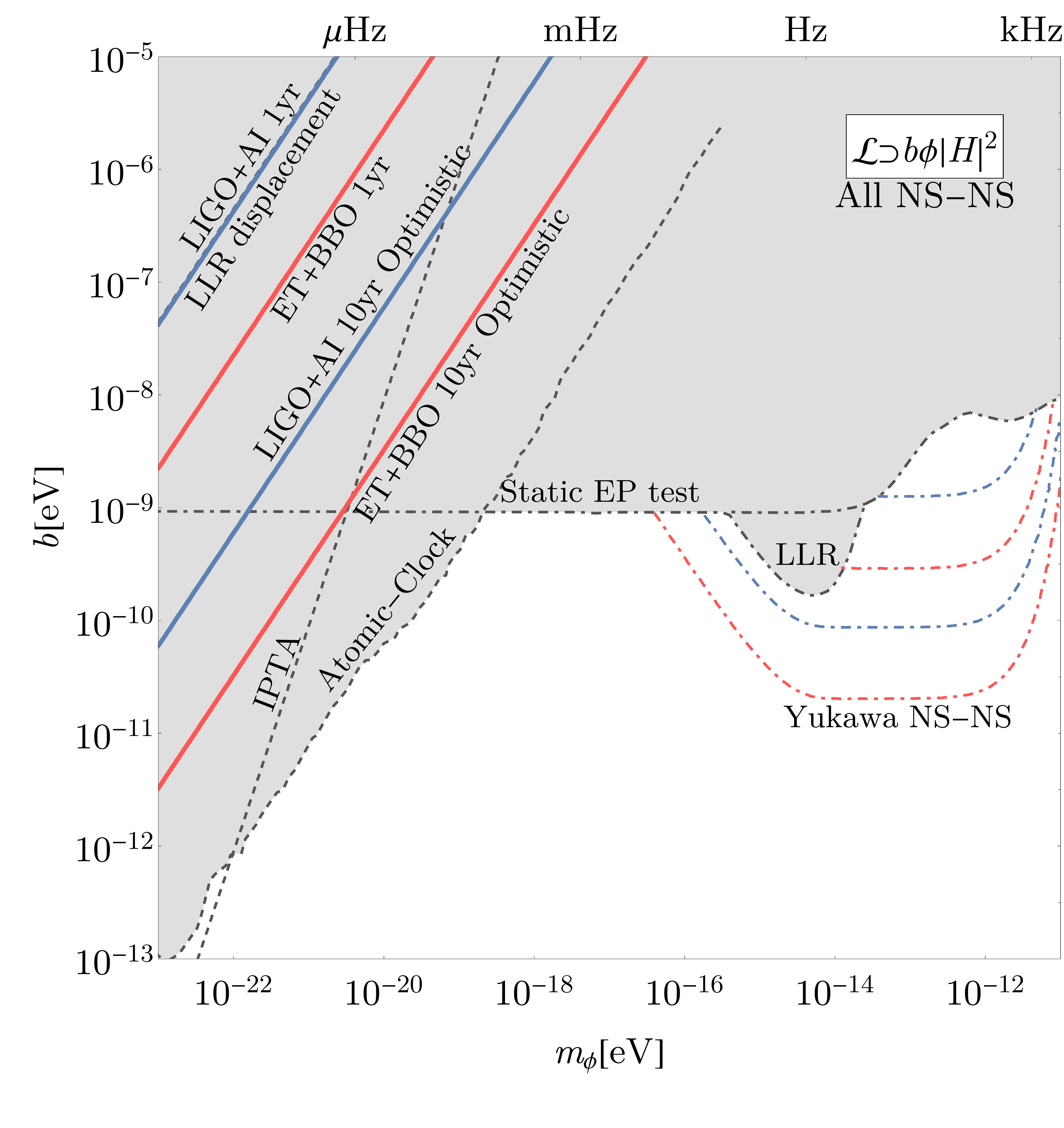}
\includegraphics[width=0.49\textwidth]{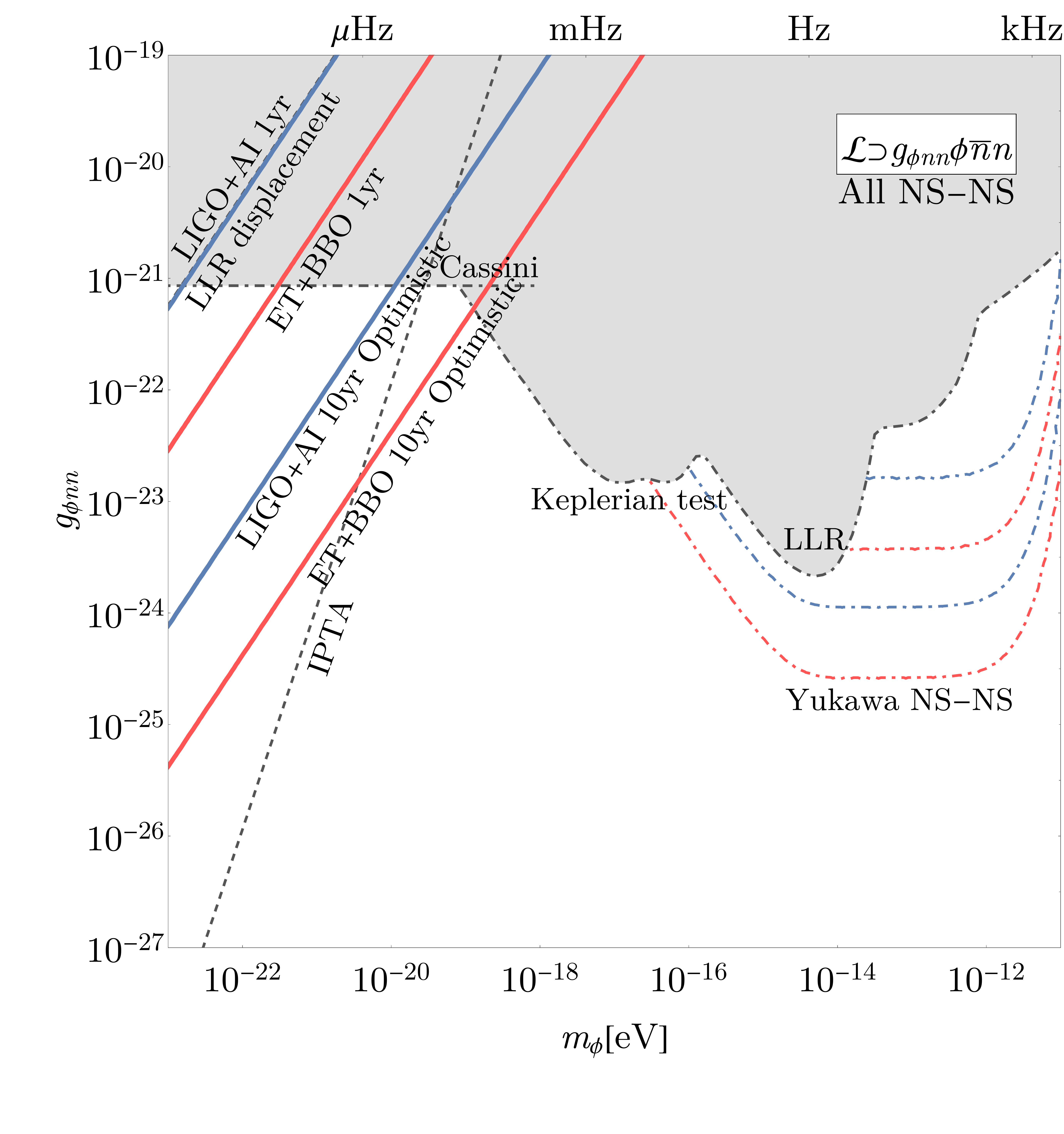}
\caption{Integration of all NS-NS measurements for 1 or 10 years with 4(aLIGO+AI) (blue) and ET+BBO (red). Along the shown solid lines, the DM-induced mass-shift $\delta \Mch/ \Mch$ equals to the integrated chirp-mass measurement accuracy. Each solid line corresponds to different setup and assumptions; see text for details. The aLIGO+AI 1yr and LLR displacement curves overlap. Models and other sensitivity curves are as in \Fig{fig:single10}.}
\label{fig:allNSNS}
\end{figure}

\Fig{fig:single10} and \Fig{fig:allNSNS} are our main results. We show the parameter space (solid lines) where the mass-shift amplitude~\Eq{eq:mass-shift} equals to the chirp-mass Fisher accuracy. \Fig{fig:single10} is from the single measurement of NS-NS at 10 Mpc, and \Fig{fig:allNSNS} is from the integration of all NS-NS measurements (for 1 or 10 years) according to their merger-rate distributions. Each solid line corresponds to different setup and assumptions, each of which will be discussed. Also shown are existing (shaded) or proposed (non-shaded) sensitivities on DM effects (dashed) from IPTA and LLR displacement and on non-DM effects (dot-dashed) from E\"{o}t-Wash EP test and Yukawa searches with LLR, Keplerian test, Cassini, and NS-NS inspiral in our experimental setups. As discussed, for the second model, we do not show weak EP-test results. And whether EP is conserved or not, IPTA probes oscillating DM effects in both models, either by pulsar mass-shift or period-variation. Notably, a large part of the light-DM parameter space of the second model is unconstrained. 

\medskip
The single measurement of NS-NS at 10 Mpc (\Fig{fig:single10}), if we are lucky to see this, alone can be already powerful. In particular, for the EP-conserving second model, this can probe unconstrained parameter space. It can also strengthen or complement other expected sensitivities from LLR displacement and IPTA. 

Each solid line shows possible improvements. The top line assumes 4(aLIGO+AI) for 1-yr integration time with full $10\times 10$ Fisher matrix, and the second line assumes a 10-yr integration with \emph{a posteriori} optical-counterpart information to remove the 5 degeneracies with sky-location ($\theta,\phi$), luminosity distance (redshift with standard cosmology), coalescence time (knowing when merges), spin $\beta$ (knowing that NS has small spin). The last two lines show ET+BBO results with the same set of assumptions. A smaller noise (LIGO+AI $\to$ ET+BBO) improves the chirp-mass accuracy by about 50 times larger SNR, whereas a longer measurement (1 year $\to$ 10 years) by about 10 times larger $N_{\rm cyc}$. Removing the 5 degeneracies improves by another factor of 2; this relatively small improvement is one of the highest-frequency benefits (see the next section). For a shorter 1-yr integration, the degeneracies with sky-location and spin $\beta$ are similar, but the latter one becomes more important for a longer 10-yr integration. This is because spin effects are irrelevant at low-frequency (farther separation) regimes, thus longer lower-frequency data do not contain much spin information.
 
 \medskip
A combination of all NS-NS observations for $T=$ 1 or 10 years of integration (\Fig{fig:allNSNS}) can extend the reach by a few orders of magnitudes. Summing all observations enhances the sensitivity statistically, scaling approximately as 
\beq
\left( \, \int dz \, \frac{4 \pi \chi(z)^2}{H(z)} \, \left( \frac{10 \, {\rm Mpc}}{\chi(z)} \right)^2 \, \frac{\dot{n}}{(1+z)^2} \, \left( \frac{ T}{\rm year} \right) \, \right)^{1/2},
\eeq
with the binary redshift $z$, comoving distance $\chi$, comoving merger-rate density of NS-NS $\dot{n}$ (we assume a constant comoving density), the Hubble constant today $H_0 = 70 \, {\rm km/sec/Mpc}$ and $\Omega_M=0.3, \, \Omega_\Lambda=0.7$ for matter and $\Lambda$ energy fraction. The two sets of predictions shown are based on lower and optimistic values of $\rho_{\rm DM}$ and $\dot{n}$. For the lower expectation, we use $\rho_{\rm DM} = 77.2 \, {\rm GeV/cm}^3$ from 0.8 kpc flat-core Burkert profile (similar to 0.1 kpc NFW and Einasto values) and $\dot{n} = 1000/{\rm Gpc}^3/{\rm yr}$~\cite{Abbott:2016ymx} from the central value of predictions. For the optimistic case, we use $\rho_{\rm DM} = 1000 \, {\rm GeV/cm}^3$~\cite{Linden:2014sra} maximum not exceeding the ${\cal O}(10)\%$ of baryonic mass inside the 100 pc galactic center (which can be constrained by future pulsar-timing residual measurements from SKA/FAST~\cite{Porayko:2018sfa}) and $\dot{n} = 12000/{\rm Gpc}^3/{\rm yr}$~\cite{Abbott:2016ymx} maximum consistent with LIGO observations so far. The ``optimistic'' curves are the optimistic results with 10-yr integration and the 5 degeneracies removed. It improves the lower sensitivity by about a factor 250. After all, the most optimistic sensitivity reaches down to an exquisite level, $b \lesssim 10^{-11}$ eV and $g_{\phi nn} \lesssim 10^{-25}$, from the 10-year integration with ET+BBO for the lightest DM.

\section{Discussion}

\subsection{Origin of good sensitivity} \label{sec:analysis}

\begin{figure}[t] \centering
\includegraphics[width=0.48\textwidth]{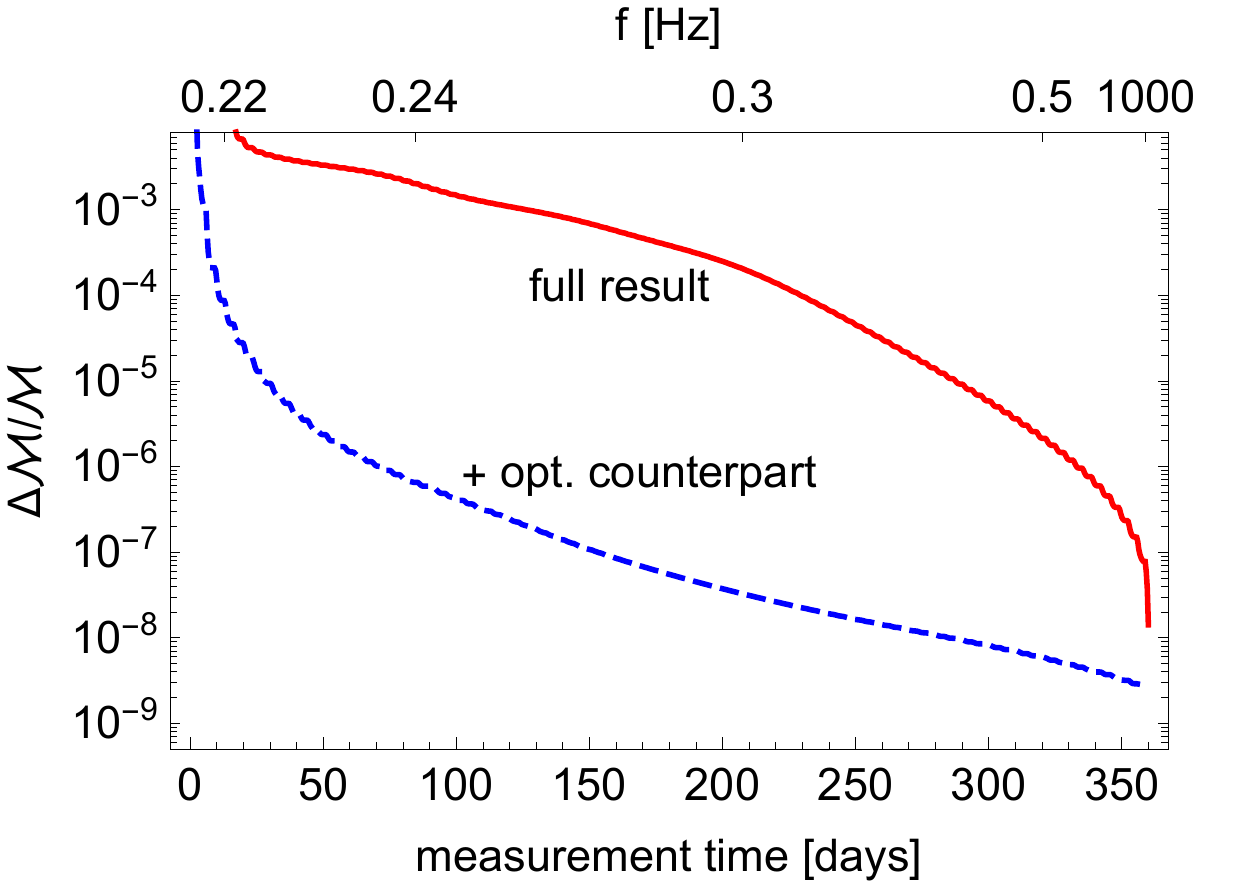}
\includegraphics[width=0.48\textwidth]{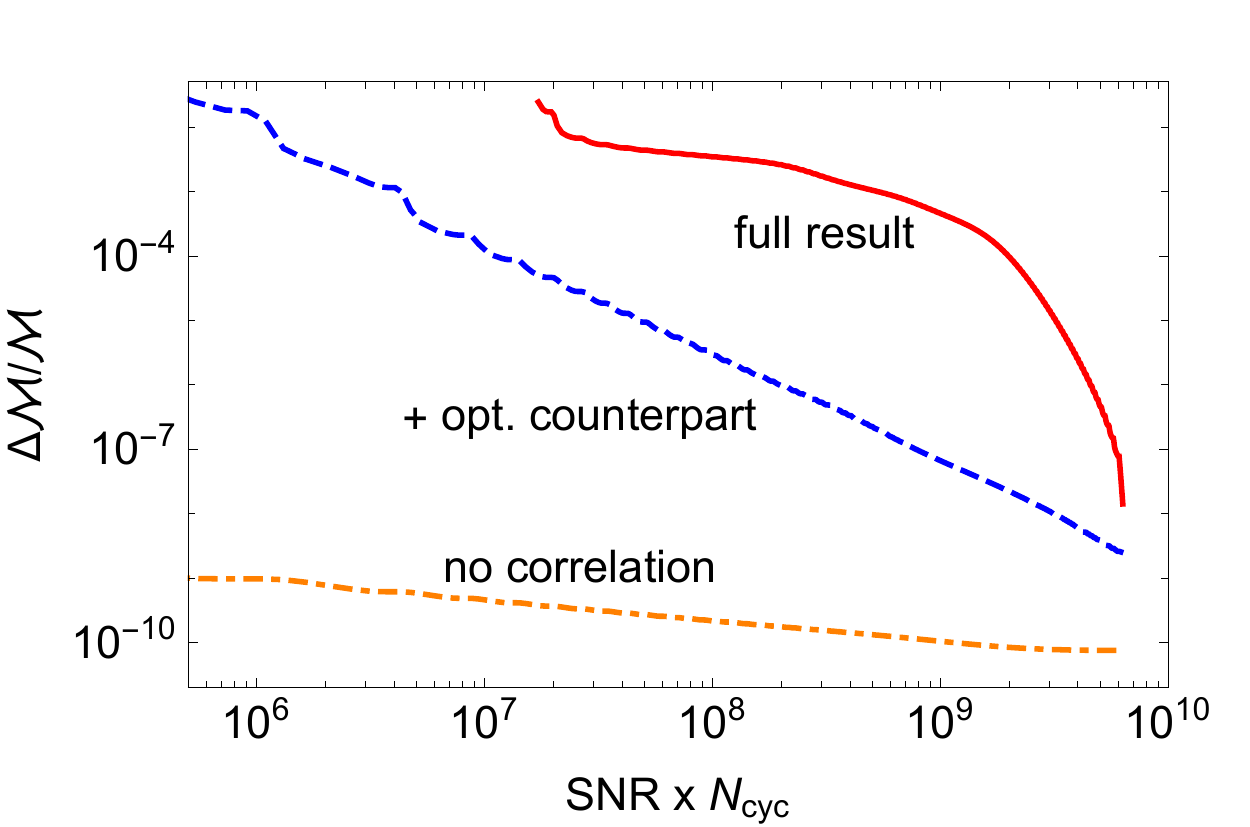}
\includegraphics[width=0.48\textwidth]{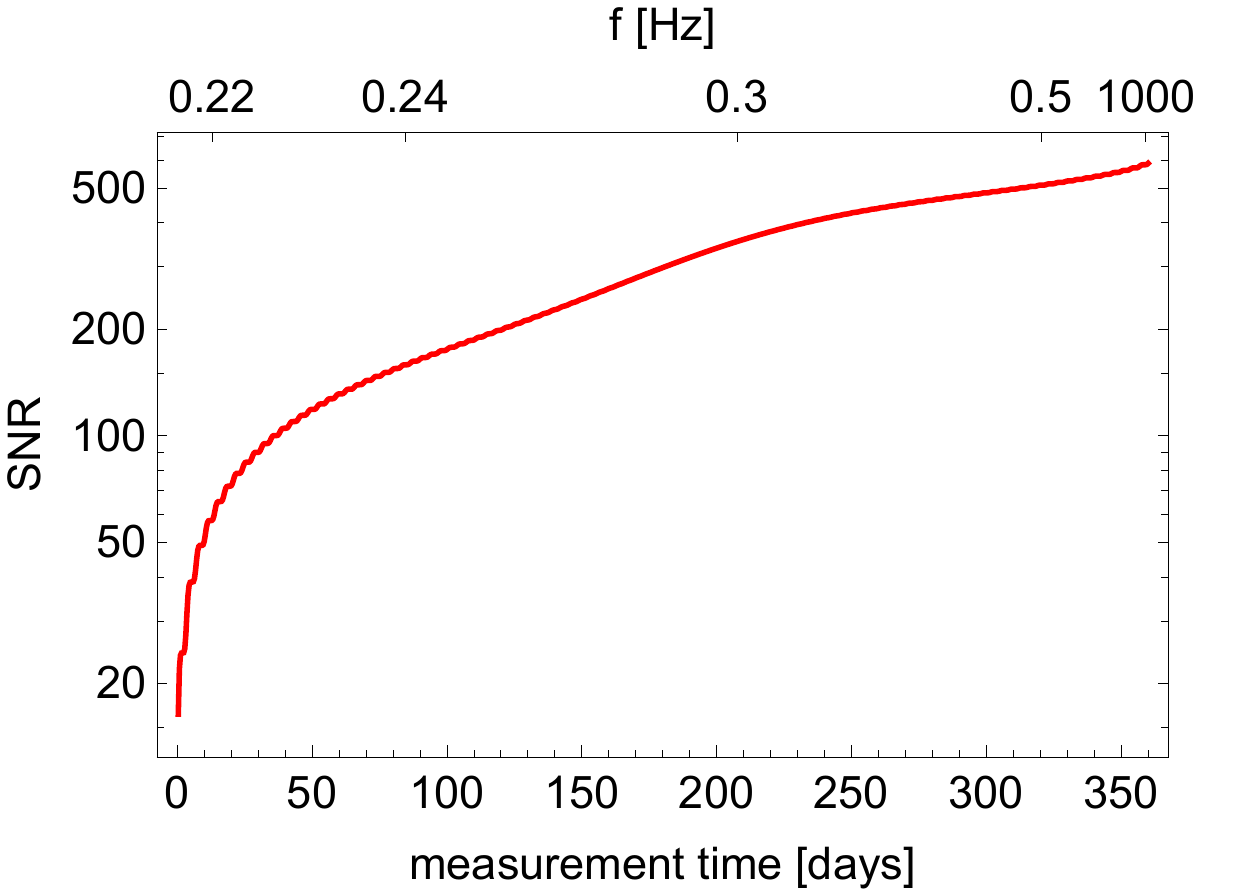}
\includegraphics[width=0.48\textwidth]{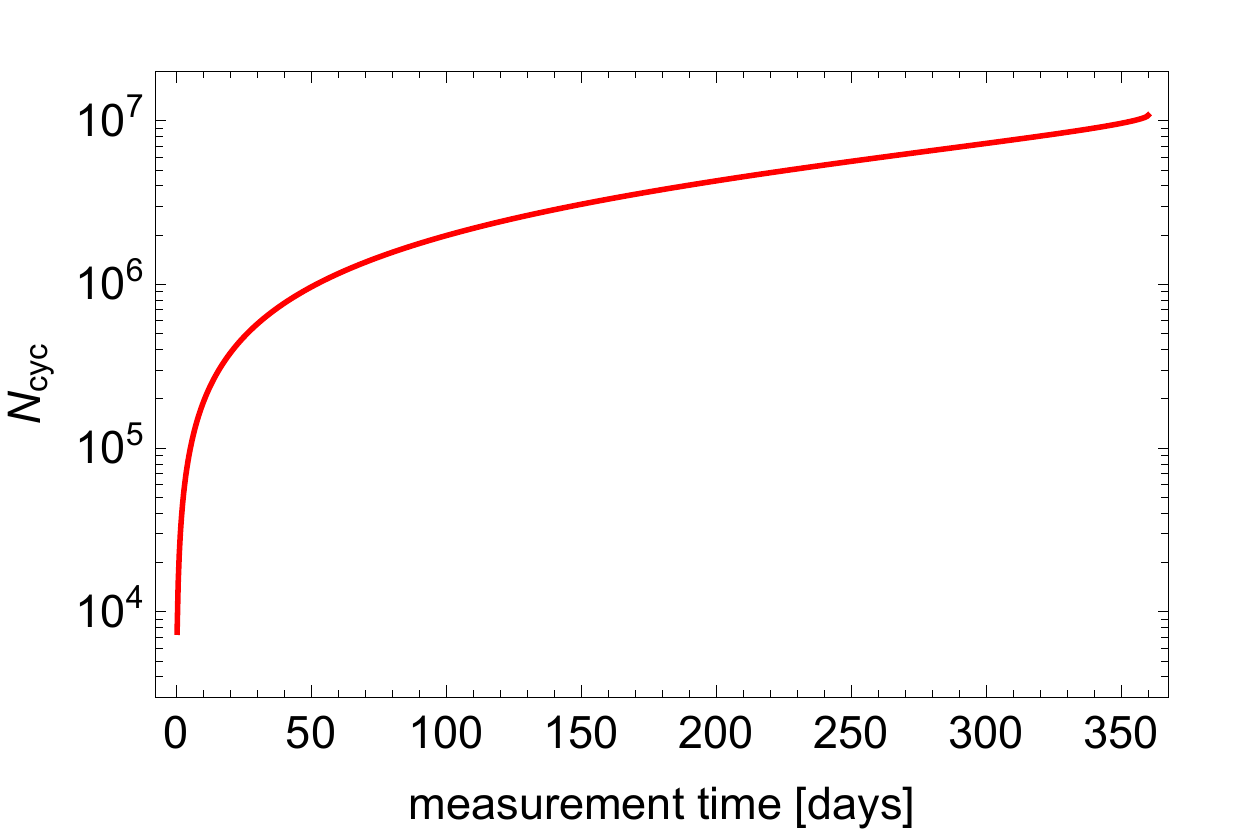}
\includegraphics[width=0.48\textwidth]{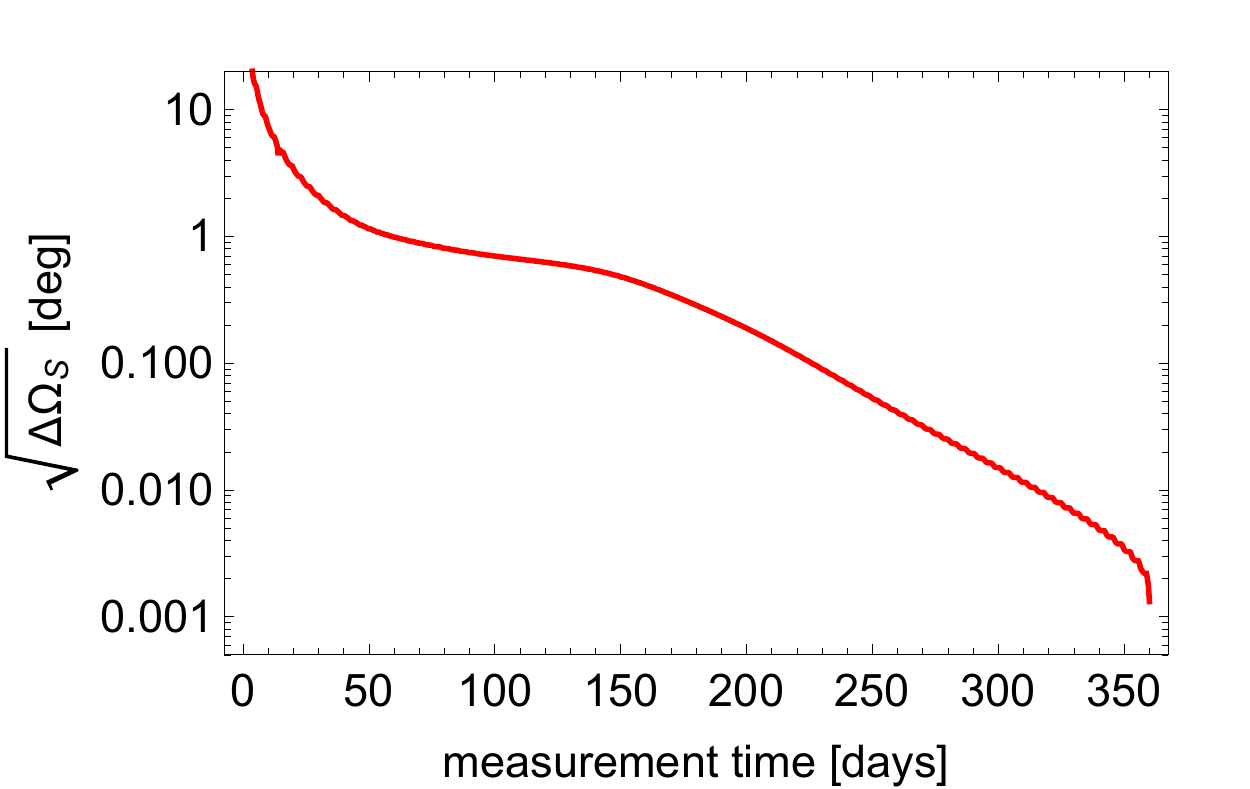}
\includegraphics[width=0.48\textwidth]{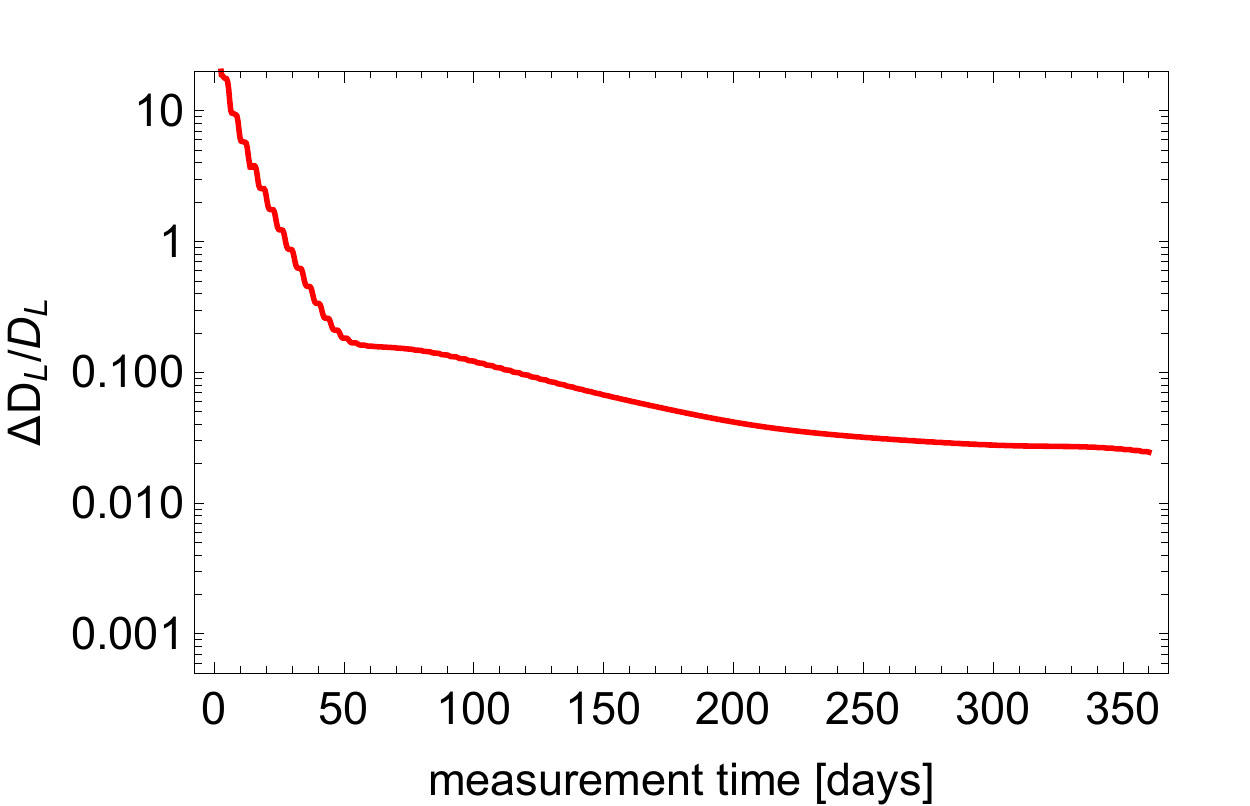}
\caption{Improvement of Fisher errors in measurement time. NS-NS at 10 Mpc. Shown parameters are chirp-mass fractional accuracy, SNR, $N_{\rm cyc}$, sky-localization accuracy, and $D_L$ fractional accuracy. The second plot shows the correlation of chirp-mass accuracy and SNR $\times N_{\rm cyc}$. The considered measurement is for the last 1 year, sweeping $f \simeq 0.22$ - 1000 Hz. The blue-dashed lines assume no correlations with the 5 source parameters (see text for details), and the orange-dotdashed line assumes no correlation with any source parameters.}
\label{fig:analysis}
\end{figure}

We now turn to analyze the origin of good sensitivity to small DM effects. 

Above all, the potential chirp-mass measurement accuracy is significantly enhanced by a large $N_{\rm cyc}$ (during year-long measurement). It is because a tiny phase shift in each cycle can be accumulated to an observably large one after $N_{\rm cyc}$ cycles~\cite{Cutler:1992tc,Cutler:1994ys}. For example, the last 1 year measurement of NS-NS at 10 Mpc yields $N_{\rm cyc} \simeq 10^7$ and SNR $\simeq 600$ (see \Fig{fig:analysis}) so that the fractional accuracy is expected to be enhanced as $1/$(SNR $\cdot \, N_{\rm cyc}) \sim 10^{-9}$, instead of just 1/SNR $\sim 10^{-2}$. Indeed, while the accuracies of parameters that do not accumulate with $N_{\rm cyc}$ (such as $\ln D_L$) is only 1/SNR $\sim 10^{-2}$, the final chirp-mass accuracy is $10^{-8}$ augmented significantly by $\sim N_{\rm cyc}$. 

As shown in the first four panels of \Fig{fig:analysis}, however, the chirp-mass accuracy does not improve closely (or linearly) with SNR $\cdot \, N_{\rm cyc}$. Only at the end of a year-long measurement, the accuracy grows significantly and becomes close to the expectation. Here, it is the interplay of low-frequency and highest-frequency regimes that allows to fully realize the potential enhancement from $N_{\rm cyc}$. We discuss this in this section.

\medskip
The relevance of $N_{\rm cyc}$ can be read directly from the Fisher matrix element. The Fisher element of the chirp-mass part, $F_{\ln \Mch \ln \Mch} \= \int df \, | d \widetilde{h} / d \ln \Mch |^2$, is given by 
\beq
\frac{d \widetilde{h}(f)}{d \ln \Mch} \, \simeq \, -\frac{5i}{4} ( 8\pi \Mch f)^{-5/3} \widetilde{h}(f), 
\label{eq:fisher-mm} \eeq
at the Newtonian order. The term in the first parenthesis is proportional to the $N_{\rm cyc}$ accumulated in each frequency interval as
\beq
N_{\rm cyc} \, \approx \, 2.44 \times 10^{7} \left( \frac{\Mch_z}{1.5 \Msun} \right)^{-5/3} \left( \frac{f_i}{10^{-1} {\rm Hz}} \right)^{-5/3}, 
\label{eq:ncyc} \eeq
where the initial frequency $f_i \ll f_f$ is assumed to be much smaller than the final frequency $f_f$. The final term $\widetilde{h}(f)$ in \Eq{eq:fisher-mm} is proportional to the SNR accumulated in that interval. Thus, the Fisher element in each frequency interval is indeed related to the $N_{\rm cyc} \cdot$ SNR there.

The most of the GW cycles is accumulated at low frequencies (see \Eq{eq:ncyc}) as GW spends much more time there than at high frequencies. So does SNR. These behaviors are shown in \Fig{fig:analysis}. Thus, a long period of particular waveform evolution at low-frequencies contains a large amount of information of the chirp mass. But a problem remains in that low-frequency data alone is not enough to confidently distinguish the chirp mass from other source parameters. The unresolved correlations with other parameters prohibit to achieve the expected accuracy yet.

Here comes the highest-frequency data, chirping rapidly towards the merger. There, the frequency evolves most rapidly, whose evolution is governed by the chirp mass. Even though $N_{\rm cyc}$ does not increase much during that short evolution, the large range of non-trivial frequency evolution can resolve various degeneracies.

In particular, high-frequency measurement significantly improves the accuracy of sky-location ($\theta, \phi$) and hence correlations with them (see \Fig{fig:analysis}). It is improved due to the huge Doppler phase shift accumulated across the Sun~\cite{Graham:2017lmg} (during year-long measurment). The GW phase explicitly grows with the frequency as
\beq
\Psi(f) \, \sim \, 2 \pi f (- \vec{r}_{AU} \cdot \hat{n} /c  + t_c  ),
\label{eq:GWphase} \eeq
where $\hat{n} = \hat{n}(\theta, \phi)$ is a unit vector for the GW propagation direction or the source's sky-location and $\vec{r}_{AU}$ is the Earth-Sun separation vector. The first term (depending on $\theta,\phi$) is the Doppler phase shift.
But this effect is measurable only after long enough measurement around the Sun~\cite{Graham:2017lmg} as linear (or constant) Doppler shift is not measurable (confused with the cosmological redshift $z$). Thus, a short high-frequency segment of data alone is not useful, but only year-long measurement can utilize this natural benefit.  As shown in \Fig{fig:analysis}, the location accuracy begins to improve  after about 6 months.

The highest-frequency chirping actually improves most of the source-parameter accuracies that affect the GW phase. The coalescence time $t_c$ will be obviously better determined as GW approaches that time (and see \Eq{eq:GWphase} that the $t_c$ contribution also grows with the frequency). Spin-orbit parameter $\beta$'s impact on precession and phase evolution will be largest when the binary separation is smallest near merger. The reduced mass $\mu$ will receive similar (although smaller) benefits that the chirp mass receives. On the other hand, the source parameters that affect GW amplitudes do not gain high-frequency benefits, as \Fig{fig:analysis} shows for $\ln D_L$ accuracy. 

By comparing the full results (red-solid) and blue-dashed lines in \Fig{fig:analysis} where we ignore the 5 aforementioned correlations (with sky-location ($\theta, \phi$), $t_c$, $D_L$, and $\beta$ from \emph{a posteriori} (optical) information), we conclude that the resolution of degeneracies is responsible for the improvement of chirp-mass accuracy at the highest-frequency end. And it is the one that finally allows to realize the potential enhancement from $N_{\rm cyc}$.

\subsection{DM density dependence} \label{sec:transparent}

The DM density dependence of the signal significance is an interesting property. If NS-NS formation follows the star formation history, then the majority of NS-NS will reside in the galactic center (GC) where DM is also most abundant. NS-NS is then a natural candidate to detect large DM effects. The DM density dependence can also be exploited to better confirm the DM origin of anomalous signals or to map DM distribution. 

In this paper, we implicitly assume that the GC is transparent to the GW. But how bright or transparent it is would be an interesting question. Note that the majority of pulsars measured (with lights) and used in PTA~\cite{Verbiest:2016vem,Porayko:2018sfa} are within about kpc from the Earth. Thus, NS-NS with the GW can also be complementary to the local fuzzy DM search by Parkes PTA~\cite{Porayko:2018sfa} too.


\section{Conclusion}

We have shown that last years of NS-NS inspiral may have a precision capability to detect tiny perturbations from the lightest possible scalar DM. The new observable considered is the time-oscillating mass shift, induced by the DM fifth force with long coherence. The precision capability stems from a large number of GW cycles and year-long measurement of highest frequencies, which can be realized by a future detector network in the broadband $f \simeq 0.01-1000$ Hz. If light scalar DM interacts with the NS, our new observable in this broadband measurement can probe a large part of the unconstrained parameter space, in particular the lightest possible mass range.

Our study also emphasizes the role of long-time high-frequency measurements in the precision GW program; a large $N_{\rm cyc}$ can enhance the chirp-mass sensitivity as discussed, and moreover, the Doppler effect around the Sun can localize the source~\cite{Graham:2017lmg,Nair:2018bxj}, and a better frequency resolution can resolve the GW lensing fringe generated by intervening cosmic strings~\cite{Jung:2018kde} and compact DM~\cite{Jung:2017flg} (see also \cite{Isoyama:2018rjb,Takahashi:2003wm} for other benefits). These shall motivate the development of mid-frequency ($f \simeq 0.01-10$ Hz) detectors that can form such broadband detector networks by combining with ongoing or upcoming LIGO-band detectors.

\begin{acknowledgments}

We thank Junwu Huang, Gungwon Kang and Dongjin Chway for useful comments. Our work is supported by National Research Foundation of Korea under grant 2015R1A4A1042542, 2017R1D1A1B03030820, by Research Settlement Fund for the new faculty of Seoul National University, and by POSCO Science Fellowship.
\end{acknowledgments}



\begin{thebibliography}{99}

\bibitem{Abbott:2016blz} 
  B.~P.~Abbott {\it et al.} [LIGO Scientific and Virgo Collaborations],
  ``Observation of Gravitational Waves from a Binary Black Hole Merger,''
  Phys.\ Rev.\ Lett.\  {\bf 116}, no. 6, 061102 (2016)
  doi:10.1103/PhysRevLett.116.061102
  [arXiv:1602.03837 [gr-qc]].
  
\bibitem{TheLIGOScientific:2016htt} 
  B.~P.~Abbott {\it et al.} [LIGO Scientific and Virgo Collaborations],
  ``Astrophysical Implications of the Binary Black-Hole Merger GW150914,''
  Astrophys.\ J.\  {\bf 818}, no. 2, L22 (2016)
  doi:10.3847/2041-8205/818/2/L22
  [arXiv:1602.03846 [astro-ph.HE]].

\bibitem{TheLIGOScientific:2016src} 
  B.~P.~Abbott {\it et al.} [LIGO Scientific and Virgo Collaborations],
  ``Tests of general relativity with GW150914,''
  Phys.\ Rev.\ Lett.\  {\bf 116}, no. 22, 221101 (2016)
  Erratum: [Phys.\ Rev.\ Lett.\  {\bf 121}, no. 12, 129902 (2018)]
  doi:10.1103/PhysRevLett.116.221101, 10.1103/PhysRevLett.121.129902
  [arXiv:1602.03841 [gr-qc]].

\bibitem{TheLIGOScientific:2017qsa} 
  B.~P.~Abbott {\it et al.} [LIGO Scientific and Virgo Collaborations],
  ``GW170817: Observation of Gravitational Waves from a Binary Neutron Star Inspiral,''
  Phys.\ Rev.\ Lett.\  {\bf 119}, no. 16, 161101 (2017)
  doi:10.1103/PhysRevLett.119.161101
  [arXiv:1710.05832 [gr-qc]].
    
\bibitem{Abbott:2017xzu} 
  B.~P.~Abbott {\it et al.} [LIGO Scientific and Virgo and 1M2H and Dark Energy Camera GW-E and DES and DLT40 and Las Cumbres Observatory and VINROUGE and MASTER Collaborations],
  ``A gravitational-wave standard siren measurement of the Hubble constant,''
  Nature {\bf 551}, no. 7678, 85 (2017)
  doi:10.1038/nature24471
  [arXiv:1710.05835 [astro-ph.CO]].
  

\bibitem{Nelson:2018xtr} 
  A.~Nelson, S.~Reddy and D.~Zhou,
  ``Dark halos around neutron stars and gravitational waves,''
  arXiv:1803.03266 [hep-ph].

\bibitem{Croon:2017zcu} 
  D.~Croon, A.~E.~Nelson, C.~Sun, D.~G.~E.~Walker and Z.~Z.~Xianyu,
  ``Hidden-Sector Spectroscopy with Gravitational Waves from Binary Neutron Stars,''
  Astrophys.\ J.\  {\bf 858}, no. 1, L2 (2018)
  doi:10.3847/2041-8213/aabe76
  [arXiv:1711.02096 [hep-ph]].

\bibitem{Ellis:2017jgp} 
  J.~Ellis, A.~Hektor, G.~Hutsi, K.~Kannike, L.~Marzola, M.~Raidal and V.~Vaskonen,
  ``Search for Dark Matter Effects on Gravitational Signals from Neutron Star Mergers,''
  Phys.\ Lett.\ B {\bf 781}, 607 (2018)
  doi:10.1016/j.physletb.2018.04.048
  [arXiv:1710.05540 [astro-ph.CO]].

\bibitem{Ellis:2018bkr} 
  J.~Ellis, G.~Hutsi, K.~Kannike, L.~Marzola, M.~Raidal and V.~Vaskonen,
  ``Dark Matter Effects On Neutron Star Properties,''
  Phys.\ Rev.\ D {\bf 97}, no. 12, 123007 (2018)
  doi:10.1103/PhysRevD.97.123007
  [arXiv:1804.01418 [astro-ph.CO]].
    
\bibitem{Randall:2017jop} 
  L.~Randall and Z.~Z.~Xianyu,
  ``Induced Ellipticity for Inspiraling Binary Systems,''
  Astrophys.\ J.\  {\bf 853}, no. 1, 93 (2018)
  doi:10.3847/1538-4357/aaa1a2
  [arXiv:1708.08569 [gr-qc]].

\bibitem{Kim:2008hd} 
  J.~E.~Kim and G.~Carosi,
  ``Axions and the Strong CP Problem,''
  Rev.\ Mod.\ Phys.\  {\bf 82}, 557 (2010)
  doi:10.1103/RevModPhys.82.557
  [arXiv:0807.3125 [hep-ph]].
  
\bibitem{Hu:2000ke} 
  W.~Hu, R.~Barkana and A.~Gruzinov,
  ``Cold and fuzzy dark matter,''
  Phys.\ Rev.\ Lett.\  {\bf 85}, 1158 (2000)
  doi:10.1103/PhysRevLett.85.1158
  [astro-ph/0003365].

\bibitem{Graham:2015cka} 
  P.~W.~Graham, D.~E.~Kaplan and S.~Rajendran,
  ``Cosmological Relaxation of the Electroweak Scale,''
  Phys.\ Rev.\ Lett.\  {\bf 115}, no. 22, 221801 (2015)
  doi:10.1103/PhysRevLett.115.221801
  [arXiv:1504.07551 [hep-ph]].
    
          
          
\bibitem{Graham:2015ifn} 
  P.~W.~Graham, D.~E.~Kaplan, J.~Mardon, S.~Rajendran and W.~A.~Terrano,
  ``Dark Matter Direct Detection with Accelerometers,''
  Phys.\ Rev.\ D {\bf 93}, no. 7, 075029 (2016)
  doi:10.1103/PhysRevD.93.075029
  [arXiv:1512.06165 [hep-ph]].

\bibitem{Piazza:2010ye} 
  F.~Piazza and M.~Pospelov,
  ``Sub-eV scalar dark matter through the super-renormalizable Higgs portal,''
  Phys.\ Rev.\ D {\bf 82}, 043533 (2010)
  doi:10.1103/PhysRevD.82.043533
  [arXiv:1003.2313 [hep-ph]].

\bibitem{Bertotti:2003rm} 
  B.~Bertotti, L.~Iess and P.~Tortora,
  Nature {\bf 425}, 374 (2003).
  doi:10.1038/nature01997

\bibitem{Freire:2012nb} 
  P.~C.~C.~Freire, M.~Kramer and N.~Wex,
  Class.\ Quant.\ Grav.\  {\bf 29}, 184007 (2012)
  doi:10.1088/0264-9381/29/18/184007
  [arXiv:1205.3751 [gr-qc]].
  
\bibitem{Archibald:2018oxs} 
  A.~M.~Archibald {\it et al.},
  Nature {\bf 559}, no. 7712, 73 (2018)
  doi:10.1038/s41586-018-0265-1
  [arXiv:1807.02059 [astro-ph.HE]].
  
\bibitem{Blas:2016ddr} 
  D.~Blas, D.~L.~Nacir and S.~Sibiryakov,
  ``Ultralight Dark Matter Resonates with Binary Pulsars,''
  Phys.\ Rev.\ Lett.\  {\bf 118}, no. 26, 261102 (2017)
  doi:10.1103/PhysRevLett.118.261102
  [arXiv:1612.06789 [hep-ph]].




  
\bibitem{Huang:2018pbu} 
  J.~Huang, M.~C.~Johnson, L.~Sagunski, M.~Sakellariadou and J.~Zhang,
  ``Prospects for axion searches with Advanced LIGO through binary mergers,''
  arXiv:1807.02133 [hep-ph].

\bibitem{Kopp:2018jom} 
  J.~Kopp, R.~Laha, T.~Opferkuch and W.~Shepherd,
  ``Cuckoo's Eggs in Neutron Stars: Can LIGO Hear Chirps from the Dark Sector?,''
  arXiv:1807.02527 [hep-ph].
  
\bibitem{Alexander:2018qzg} 
  S.~Alexander, E.~McDonough, R.~Sims and N.~Yunes,
  ``Hidden-Sector Modifications to Gravitational Waves From Binary Inspirals,''
  arXiv:1808.05286 [gr-qc].
      
\bibitem{Arvanitaki:2015iga} 
  A.~Arvanitaki, S.~Dimopoulos and K.~Van Tilburg,
  ``Sound of Dark Matter: Searching for Light Scalars with Resonant-Mass Detectors,''
  Phys.\ Rev.\ Lett.\  {\bf 116}, no. 3, 031102 (2016)
  doi:10.1103/PhysRevLett.116.031102
  [arXiv:1508.01798 [hep-ph]].
  
\bibitem{VanTilburg:2015oza} 
  K.~Van Tilburg, N.~Leefer, L.~Bougas and D.~Budker,
  ``Search for ultralight scalar dark matter with atomic spectroscopy,''
  Phys.\ Rev.\ Lett.\  {\bf 115}, no. 1, 011802 (2015)
  doi:10.1103/PhysRevLett.115.011802
  [arXiv:1503.06886 [physics.atom-ph]].
  
\bibitem{Arvanitaki:2014faa} 
  A.~Arvanitaki, J.~Huang and K.~Van Tilburg,
  ``Searching for dilaton dark matter with atomic clocks,''
  Phys.\ Rev.\ D {\bf 91}, no. 1, 015015 (2015)
  doi:10.1103/PhysRevD.91.015015
  [arXiv:1405.2925 [hep-ph]].

\bibitem{Stadnik:2015xbn} 
  Y.~V.~Stadnik and V.~V.~Flambaum,
  ``Enhanced effects of variation of the fundamental constants in laser interferometers and application to dark matter detection,''
  Phys.\ Rev.\ A {\bf 93}, no. 6, 063630 (2016)
  doi:10.1103/PhysRevA.93.063630
  [arXiv:1511.00447 [physics.atom-ph]].

\bibitem{Stadnik:2014tta} 
  Y.~V.~Stadnik and V.~V.~Flambaum,
  ``Searching for dark matter and variation of fundamental constants with laser and maser interferometry,''
  Phys.\ Rev.\ Lett.\  {\bf 114}, 161301 (2015)
  doi:10.1103/PhysRevLett.114.161301
  [arXiv:1412.7801 [hep-ph]].
            
\bibitem{Verbiest:2016vem} 
  J.~P.~W.~Verbiest {\it et al.},
  ``The International Pulsar Timing Array: First Data Release,''
  Mon.\ Not.\ Roy.\ Astron.\ Soc.\  {\bf 458}, no. 2, 1267 (2016)
  doi:10.1093/mnras/stw347
  [arXiv:1602.03640 [astro-ph.IM]].

\bibitem{Porayko:2018sfa} 
  N.~K.~Porayko {\it et al.},
  Phys.\ Rev.\ D {\bf 98}, no. 10, 102002 (2018)
  doi:10.1103/PhysRevD.98.102002
  [arXiv:1810.03227 [astro-ph.CO]].

\bibitem{Williams:2012nc} 
  J.~G.~Williams, S.~G.~Turyshev and D.~Boggs,
  ``Lunar Laser Ranging Tests of the Equivalence Principle,''
  Class.\ Quant.\ Grav.\  {\bf 29}, 184004 (2012)
  doi:10.1088/0264-9381/29/18/184004
  [arXiv:1203.2150 [gr-qc]].
  

\bibitem{Arvanitaki:2016fyj} 
  A.~Arvanitaki, P.~W.~Graham, J.~M.~Hogan, S.~Rajendran and K.~Van Tilburg,
  ``Search for light scalar dark matter with atomic gravitational wave detectors,''
  Phys.\ Rev.\ D {\bf 97}, no. 7, 075020 (2018)
  doi:10.1103/PhysRevD.97.075020
  [arXiv:1606.04541 [hep-ph]].

\bibitem{Pierce:2018xmy} 
  A.~Pierce, K.~Riles and Y.~Zhao,
  ``Searching for Dark Photon Dark Matter with Gravitational Wave Detectors,''
  Phys.\ Rev.\ Lett.\  {\bf 121}, no. 6, 061102 (2018)
  doi:10.1103/PhysRevLett.121.061102
  [arXiv:1801.10161 [hep-ph]].

\bibitem{Adelberger:2003zx} 
  E.~G.~Adelberger, B.~R.~Heckel and A.~E.~Nelson,
  ``Tests of the gravitational inverse square law,''
  Ann.\ Rev.\ Nucl.\ Part.\ Sci.\  {\bf 53}, 77 (2003)
  doi:10.1146/annurev.nucl.53.041002.110503
  [hep-ph/0307284].

\bibitem{Wagner:2012ui} 
  T.~A.~Wagner, S.~Schlamminger, J.~H.~Gundlach and E.~G.~Adelberger,
  ``Torsion-balance tests of the weak equivalence principle,''
  Class.\ Quant.\ Grav.\  {\bf 29}, 184002 (2012)
  doi:10.1088/0264-9381/29/18/184002
  [arXiv:1207.2442 [gr-qc]].
  
\bibitem{Touboul:2017grn} 
  P.~Touboul {\it et al.},
  ``MICROSCOPE Mission: First Results of a Space Test of the Equivalence Principle,''
  Phys.\ Rev.\ Lett.\  {\bf 119}, no. 23, 231101 (2017)
  doi:10.1103/PhysRevLett.119.231101
  [arXiv:1712.01176 [astro-ph.IM]].
  
\bibitem{Hees:2018fpg} 
  A.~Hees, O.~Minazzoli, E.~Savalle, Y.~V.~Stadnik and P.~Wolf,
  ``Violation of the equivalence principle from light scalar dark matter,''
  Phys.\ Rev.\ D {\bf 98}, no. 6, 064051 (2018)
  doi:10.1103/PhysRevD.98.064051
  [arXiv:1807.04512 [gr-qc]].
 
\bibitem{Hees:2016gop} 
  A.~Hees, J.~Guena, M.~Abgrall, S.~Bize and P.~Wolf,
  ``Searching for an oscillating massive scalar field as a dark matter candidate using atomic hyperfine frequency comparisons,''
  Phys.\ Rev.\ Lett.\  {\bf 117}, no. 6, 061301 (2016)
  doi:10.1103/PhysRevLett.117.061301
  [arXiv:1604.08514 [gr-qc]].
   
\bibitem{Stadnik:2016zkf} 
  Y.~V.~Stadnik and V.~V.~Flambaum,
  ``Improved limits on interactions of low-mass spin-0 dark matter from atomic clock spectroscopy,''
  Phys.\ Rev.\ A {\bf 94}, no. 2, 022111 (2016)
  doi:10.1103/PhysRevA.94.022111
  [arXiv:1605.04028 [physics.atom-ph]].


\bibitem{TheLIGOScientific:2016agk} 
  B.~P.~Abbott {\it et al.} [LIGO Scientific and Virgo Collaborations],
  ``GW150914: The Advanced LIGO Detectors in the Era of First Discoveries,''
  Phys.\ Rev.\ Lett.\  {\bf 116}, no. 13, 131103 (2016)
  doi:10.1103/PhysRevLett.116.131103
  [arXiv:1602.03838 [gr-qc]].

\bibitem{Graham:2016plp} 
  P.~W.~Graham, J.~M.~Hogan, M.~A.~Kasevich and S.~Rajendran,
  ``Resonant mode for gravitational wave detectors based on atom interferometry,''
  Phys.\ Rev.\ D {\bf 94}, no. 10, 104022 (2016)
  doi:10.1103/PhysRevD.94.104022
  [arXiv:1606.01860 [physics.atom-ph]].
  
\bibitem{Graham:2017pmn} 
  P.~W.~Graham {\it et al.} [MAGIS Collaboration],
  ``Mid-band gravitational wave detection with precision atomic sensors,''
  arXiv:1711.02225 [astro-ph.IM].

\bibitem{Hild:2010id} 
  S.~Hild {\it et al.},
  ``Sensitivity Studies for Third-Generation Gravitational Wave Observatories,''
  Class.\ Quant.\ Grav.\  {\bf 28}, 094013 (2011)
  doi:10.1088/0264-9381/28/9/094013
  [arXiv:1012.0908 [gr-qc]].

\bibitem{Yagi:2011wg} 
  K.~Yagi and N.~Seto,
  ``Detector configuration of DECIGO/BBO and identification of cosmological neutron-star binaries,''
  Phys.\ Rev.\ D {\bf 83}, 044011 (2011)
  Erratum: [Phys.\ Rev.\ D {\bf 95}, no. 10, 109901 (2017)]
  doi:10.1103/PhysRevD.95.109901, 10.1103/PhysRevD.83.044011
  [arXiv:1101.3940 [astro-ph.CO]].



\bibitem{Graham:2017lmg} 
  P.~W.~Graham and S.~Jung,
  ``Localizing Gravitational Wave Sources with Single-Baseline Atom Interferometers,''
  Phys.\ Rev.\ D {\bf 97}, no. 2, 024052 (2018)
  doi:10.1103/PhysRevD.97.024052
  [arXiv:1710.03269 [gr-qc]].

\bibitem{Cutler:1997ta} 
  C.~Cutler,
  ``Angular resolution of the LISA gravitational wave detector,''
  Phys.\ Rev.\ D {\bf 57}, 7089 (1998)
  doi:10.1103/PhysRevD.57.7089
  [gr-qc/9703068].

  
\bibitem{Abbott:2016ymx} 
  B.~P.~Abbott {\it et al.} [LIGO Scientific and Virgo Collaborations],
  ``Upper Limits on the Rates of Binary Neutron Star and Neutron Star-black Hole Mergers From Advanced Ligo's First Observing run,''
  Astrophys.\ J.\  {\bf 832}, no. 2, L21 (2016)
  doi:10.3847/2041-8205/832/2/L21
  [arXiv:1607.07456 [astro-ph.HE]].

\bibitem{Linden:2014sra} 
  T.~Linden,
  ``Dark matter in the Galactic center,''
  IAU Symp.\  {\bf 303}, 403 (2014).
  doi:10.1017/S1743921314001021



\bibitem{Cutler:1992tc} 
  C.~Cutler {\it et al.},
  ``The Last three minutes: issues in gravitational wave measurements of coalescing compact binaries,''
  Phys.\ Rev.\ Lett.\  {\bf 70}, 2984 (1993)
  doi:10.1103/PhysRevLett.70.2984
  [astro-ph/9208005].

\bibitem{Cutler:1994ys} 
  C.~Cutler and E.~E.~Flanagan,
  ``Gravitational waves from merging compact binaries: How accurately can one extract the binary's parameters from the inspiral wave form?,''
  Phys.\ Rev.\ D {\bf 49}, 2658 (1994)
  doi:10.1103/PhysRevD.49.2658
  [gr-qc/9402014].


\bibitem{Nair:2018bxj} 
  R.~Nair and T.~Tanaka,
  ``Synergy between ground and space based gravitational wave detectors. Part II: Localisation,''
  JCAP {\bf 1808}, no. 08, 033 (2018)
  doi:10.1088/1475-7516/2018/08/033
  [arXiv:1805.08070 [gr-qc]].

\bibitem{Jung:2018kde} 
  S.~Jung and T.~Kim,
  ``A New Probe of Cosmic Strings at LIGO and Mid-band: Gravitational Lensing Fringe,''
  arXiv:1810.04172 [astro-ph.CO].
  
\bibitem{Jung:2017flg} 
  S.~Jung and C.~S.~Shin,
  ``Gravitational-Wave Lensing Fringes by Compact Dark Matter at LIGO,''
  arXiv:1712.01396 [astro-ph.CO].

\bibitem{Isoyama:2018rjb} 
  S.~Isoyama, H.~Nakano and T.~Nakamura,
  ``Multiband Gravitational-Wave Astronomy: Observing binary inspirals with a decihertz detector, B-DECIGO,''
  PTEP {\bf 2018}, no. 7, 073E01 (2018)
  doi:10.1093/ptep/pty078
  [arXiv:1802.06977 [gr-qc]].

\bibitem{Takahashi:2003wm} 
  R.~Takahashi and T.~Nakamura,
  ``Deci hertz laser interferometer can determine the position of the coalescing binary neutron stars within an arc minute a week before the final merging event to black hole,''
  Astrophys.\ J.\  {\bf 596}, L231 (2003)
  doi:10.1086/379112
  [astro-ph/0307390].

                                                          
\end{thebibliography}
\end{document}